# Doppler-free spectroscopy of molecular iodine using a frequency-stable light source at 578 nm

Feng-Lei Hong*, Hajime Inaba, Kazumoto Hosaka, Masami Yasuda, and Atsushi Onae

*National Metrology Institute of Japan (NMIJ), National Institute of Advanced Industrial Science and Technology (AIST), Tsukuba Central 3, Ibaraki 305-8563, Japan*
*CREST, Japan Science and Technology Agency, 4-1-8 Honcho Kawaguchi, Saitama 332-0012, Japan*
*\*Corresponding author: f.hong@aist.go.jp*

**Abstract:** A stable light source obtained using sum-frequency generation (SFG) is developed for high-resolution spectroscopy at 578 nm. Hyperfine transitions of molecular iodine are observed by using the SFG light source with saturation spectroscopy. The light source is frequency stabilized to the observed hyperfine transition and achieves a stability of $2\times10^{-12}$ for a 1-s averaging time. The absolute frequency of the light source stabilized on the $a_1$ component of the $R(37)16$-$1$ transition is determined as 518 304 551 833 (2) kHz. This transition serves as a frequency reference for the $^1S_0 - {}^3P_0$ optical clock transition in neutral ytterbium (Yb).

## 1. Introduction

Coherent light sources in the visible region are important for a number of scientific and technical applications including metrology, remote sensing, communications and biotechnology. It is especially important to find stable light sources for high-resolution spectroscopy and metrology applications. Laser-diode (LD)-pumped solid-state lasers and fiber lasers have excellent frequency characteristics but are only available at certain limited wavelengths around 1-1.5 μm. LD frequency is easy to control and LDs can cover wavelengths of around 400 nm and longer than 600 nm. However, some interesting spectroscopic targets, for example the Yb [1] and Hg⁺ [2] systems, employ transitions at 500 – 600 nm. Molecular iodine also has a series of absorption lines in this wavelength region [3]. Dye lasers are usually used in such systems. However dye lasers are difficult to handle and considerable effort is needed to narrow the linewidth, therefore we need to establish an alternative light source.

At CLEO 2006, we proposed a sum-frequency generation (SFG) scheme for the clock transition of Yb at 578 nm [4]. We used a 1319-nm Nd:YAG laser and a 1030-nm Yb-doped fiber laser to generate the SFG light. A similar idea was also adopted by a group at the National Institute of Standards and Technology (NIST) for the Yb clock transition [5]. Recently, a diode laser system was developed at 1156 nm that provides a second-harmonic-generation (SHG) power of more than 3 mW at 578 nm [6]. With the SFG scheme, an initial linewidth of ~ kHz can be easily obtained by using monolithic-cavity solid-state lasers [7] (a 1319-nm Nd:YAG laser and a 1030-nm Yb:YAG laser [8]), but with the cost of preparing two lasers. On the other hand, with the SHG scheme, the initial linewidth of a diode laser system is usually ~ 100 kHz, but with a relatively low cost and a fast current control port. Recent excellent results have revealed the potential for using an Yb lattice clock as a next generation optical frequency standard [9-12].

The hyperfine spectrum of iodine is important because many recommended optical frequency standards are based on the hyperfine transitions of molecular iodine [13]. Iodine-stabilized Nd:YAG lasers have attracted considerable attention owing to their excellent frequency stability and reproducibility [14-16]. Hyperfine transitions of molecular iodine near 532 nm have been studied by several groups with improving accuracy [17-19]. Hyperfine transitions at 660 nm have been studied by using a frequency-doubled Nd:YAG laser at 1319 nm [20]. As indicated in Fig. 1, there are three iodine lines near the $^1S_0 - {}^3P_0$ transition of Yb at 578 nm [3]. The hyperfine transitions of these lines could be important frequency reference lines for Yb atom research. As yet there have been no reports on these hyperfine transitions.

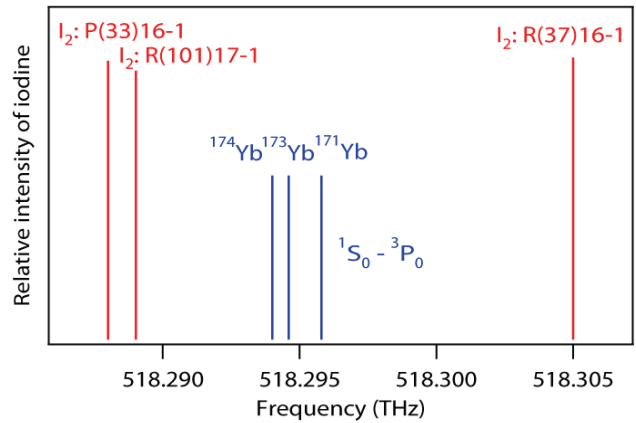

Fig. 1. Frequency atlas of the $^{127}I_2$ transitions near the $^1S_0 - {}^3P_0$ transition of Yb at 578 nm. The relative intensity of iodine is taken from Ref. [3].

In this paper, we demonstrate the Doppler-free spectroscopy of molecular iodine using an SFG light source at 578 nm. A high conversion efficiency is achieved for SFG with an MgO-doped periodically-poled-lithium-niobate (PPLN) waveguide (WG) device. Saturation spectroscopy is used to observe the Doppler-free spectra of the iodine lines. Frequency stabilization is demonstrated with the SFG light source. The frequency stability is evaluated and the absolute frequency is measured by using an optical frequency comb based on a mode-locked Ti:sapphire laser, which is phase locked to a hydrogen-maser (H-maser). The absolute frequency of the $a_1$ component of the $R(37)16-1$ transition is found to be 518304551833 (2) kHz. Since the $^1S_0 - {}^3P_0$ optical clock transition in neutral Yb is very weak, a nearby frequency reference line is particularly helpful, especially for atomic physics laboratories that have no optical frequency comb.

## 2. Experimental setup

Figure 2 is a schematic diagram of the experimental setup. The 1319-nm Nd:YAG laser with a monolithic ring cavity has a linewidth of ~ kHz. The laser emits about 900 mW of power in free space. The 1030-nm Yb DFB fiber laser has a linewidth of ~ 10 kHz. The laser emits about 200 mW of power in an optical fiber. The 1319-nm light was coupled into a polarization-maintaining (PM) fiber and was mixed with the 1030-nm light by using a PM wavelength-division multiplexing (WDM) coupler. The PPLN-WG device was designed to have a PM fiber pigtail so that the output from the WDM fiber coupler is directly connected to the fiber pigtail thus realizing stable long-term operation. The generated 578-nm SFG light was collimated by using an objective lens and was separated out from the original 1319-nm and 1030-nm lights by using a pair of dichroic mirrors.

A λ/2 plate was used to rotate the beam polarization plane such that a polarization beam splitter (PBS1) divided the 578-nm beam with an appropriate ratio into an optical frequency comb and an iodine spectrometer. The frequency comb was based on a mode-locked Ti:sapphire laser, which was self-referenced and phase locked to an H-maser. We have previously used Ti:sapphire combs to measure the absolute frequencies of an iodine-stabilized Nd:YAG laser, an iodine-stabilized He-Ne laser, an acetylene-stabilized laser, and a Sr optical lattice clock [21-24].

The Doppler-free spectroscopy of molecular iodine was based on the modulation transfer technique [14] of saturation spectroscopy. A half-wave plate and another polarization beam splitter (PBS2) were introduced to adjust the power ratio of the pump and probe beams in the spectrometer. The pump beam was frequency shifted by an acousto-optic modulator and was phase modulated by an electro-optic modulator. The acousto-optic modulator worked as an optical isolator to prevent interferometric baseline problems in the iodine spectrometer. The pump and probe beams were overlapped to provide counterpropagating beams within a 45-cm-long iodine cell. The unmodulated probe beam passed through the iodine cell and developed sidebands as a result of nonlinear resonant four-wave mixing when saturation occurred. This probe beam was steered by a third polarization beam splitter (PBS3) onto a photo detector. We obtained the modulation transfer signal of the spectral lines by demodulating the signal from the detector. The servo signal was fed back to the piezoelectric transducer actuator of the 1030-nm Yb fiber laser.

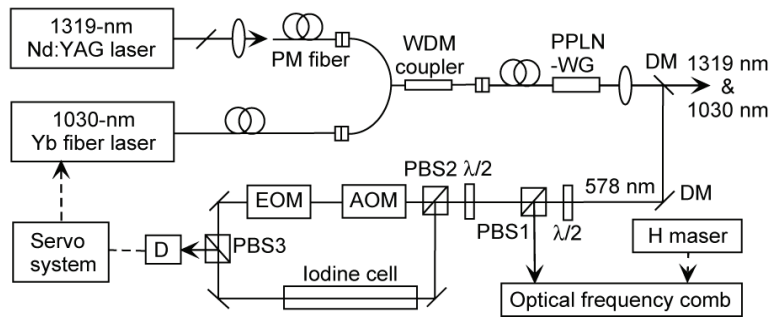

Fig. 2. Schematic diagram of the experimental setup. PM fiber, polarization-maintaining fiber; WDM, wavelength-division multiplexing; PPLN, periodically poled lithium niobate; WG, waveguide; DM, dichroic mirror; PBS1, PBS2, PBS3, polarization beam splitters; AOM, acousto-optic modulator; EOM, electro-optic modulator; D, detector.

## 3. Experimental results

### 3.1 Sum-frequency generation

Figure 3 shows the measured relation between the input power of the 1030-nm Yb fiber laser and the SFG output power at 578 nm, when the input power of the 1319-nm Nd: YAG laser was 40 mW. The indicated input powers of the two IR lasers were measured separately after the PPLN-WG (when there was no SFG). The SFG power at 578 nm was measured after the PPLN-WG with a distinction ratio between the SFG and the IR lights of more than $10^{-3}$ using a pair of dichroic mirrors. An SFG power of 22 mW was obtained using the 1319-nm and 1030-nm lasers with input powers of 40 mW and 25 mW, respectively. The PPLN-WG was 20 mm long. This corresponds to a normalized conversion coefficient of 550 %$W^{-1}cm^{-2}$. The inset in Fig. 3 shows the measured phase-matching curve of the SFG when the temperature of the crystal was varied. The observed full width of the phase-matching curve was about 2 ºC. This means no critical temperature control is needed for the PPLN-WG device.

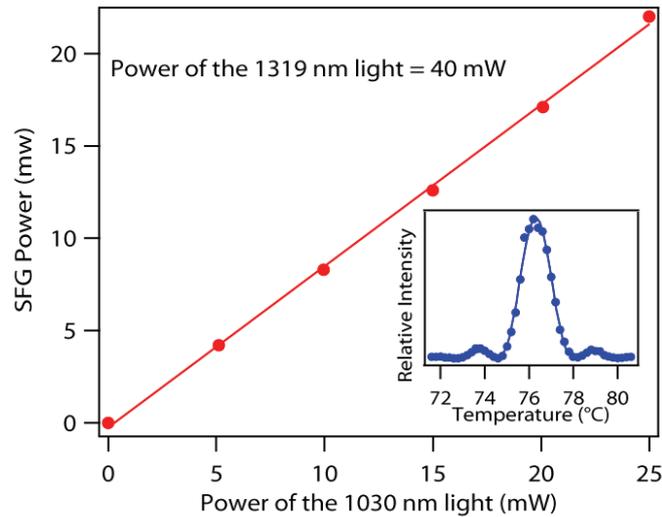

Fig. 3. Relation between the input power of the 1030-nm Yb fiber laser and the SFG output power at 578 nm, when the input power of the 1319-nm Nd: YAG laser was 40 mW. The inset shows the phase-matching curve of the SFG when the temperature of the crystal was varied.

*3.2 Saturation spectroscopy of molecular iodine*

Figure 4 shows the observed hyperfine transitions of the *R*(37)16-1 line. In this case, the frequency of the 1319-nm Nd:YAG laser was stabilized to an iodine line at 660 nm using its second harmonics [20]. We obtained a continuous frequency scan across the hyperfine transitions of the *R*(37)16-1 line at 578 nm by tuning the PZT of the 1030-nm Yb fiber laser. The optical powers of the pump and probe beams in the iodine spectrometer were 2.1 and 0.19 mW, respectively. The diameters of collimated beams inside the cell were ~ 1.5 mm. The cold-finger temperature of the iodine cell was held at – 5 ºC, corresponding to an iodine pressure of 2.4 Pa. The temperature of the cell body matched the controlled room temperature of 22 ºC. The modulation frequency of the electro-optic modulator was 354 kHz. The pump beam was frequency shifted by 80 MHz by the acousto-optic modulator. For an odd number *J* of ground rotational states (in this case $J'' = 37$), the rotational–vibrational energy level was split into 21 sublevels, resulting in 21 hyperfine components. In the observed *R*(37)16-1 transition, all the 21 hyperfine components are isolated from each other. The inset in Fig. 4 shows a horizontally expanded signal of the $a_4$ component. The signal-to-noise ratio (S/N) of the $a_4$ component was approximately 130 in a bandwidth of 10 Hz. The linewidth of the $a_4$ component was ~ 760 kHz. The flat baseline of the observed spectra in the present experiment is important as regards minimizing the frequency instability.

From the observed smooth hyperfine line, we can confirm the small frequency jitter and linewidth of the SFG light source. The hyperfine structure could not be fully resolved with a usual dye laser.

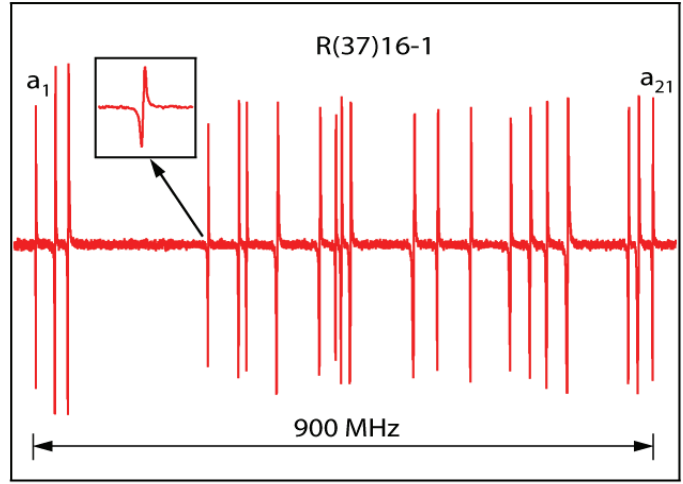

Fig. 4. Doppler-free spectra of the *R*(37)16-1 transition of molecular iodine. All 21 hyperfine components are isolated. The inset shows the horizontally expanded signal of the $a_4$ hyperfine component.

*3.3 Frequency stabilization and absolute frequency measurement*

The frequency of the SFG light source is stabilized by servo-controlling the 1030-nm Yb fiber laser using the observed hyperfine transitions. In this case, the 1319-nm Nd:YAG laser can be either frequency stabilized or free running. The frequency stability of the SFG light source was measured by using an optical frequency comb. In Fig. 5, the solid curve with solid circles indicates the root Allan variance (Allan deviation) of the measured beat frequency between the SFG light locked on the $R(37)16\text{-}1\text{:}a_1$ transition and the comb locked on the H-maser. The beat frequency was measured by using a frequency counter (Agilent 53132A) with an internal filter function. The Allan variance [25] is calculated as follows,

$$\sigma^2(\tau) = \frac{1}{2\nu^2(M-1)} \sum_{i=1}^{M-1}(y_{i+1} - y_i)^2$$

where $\tau$ is the time between successive measurements as well as the duration of each frequency measurement, $\nu$ is the mean optical frequency, $M$ is the number of measurements, and $y_i$ is the $i$th frequency measurement.

Since the stability of the H-maser-based comb (shown as a dashed curve in Fig. 5) was better than that of the iodine-stabilized SFG light source, the observed stability of the measured beat frequency was mainly limited by the frequency stability of the SFG light source. The frequency stability of the SFG laser was $2\times10^{-12}$ for a 1-s averaging time, improving toward the $7\times10^{-13}$ level after 10 s. At an averaging time > 10 s, the Allan deviation of the SFG laser appears to reach the flicker floor of the system. For comparison, the Allan deviations of iodine-stabilized Nd:YAG lasers at 532 nm and 660 nm are also shown in Fig. 5 by the dotted and solid lines, respectively. At a 1-s averaging time, the laser stability is basically determined by the S/N and the linewidth of the observed transition. Stronger lines [3] with a narrower linewidth [26] have been observed when the wavelength is closer to 500 nm, which is the dissociation limit of molecular iodine. This means that a frequency-stabilized laser using an iodine transition closer to 500 nm has better short-term frequency stability. The observed 1-s frequency stability of our iodine-stabilized lasers at different wavelengths follows this trend.

We performed absolute frequency measurements with the iodine-stabilized SFG light source. The laser was locked on the $a_1$ component of the $R(37)16\text{-}1$ transition. The $R(37)16\text{-}1$ transition was selected because it is close to the $^1S_0 - {}^3P_0$ optical clock transition of Yb and is well isolated from other iodine lines (as shown Fig. 1). The $a_1$ hyperfine component was selected because it is the closest to the Yb optical clock transition of all the hyperfine components of the $R(37)16-1$ transition. The measured frequency of the $R(37)16-1:a_1$ hyperfine transition is 518304551833 kHz. This was calculated from 1000 pieces of beat frequency data, where each piece was measured with a 1-s gate time by a frequency counter. The frequency reproducibility due to the process of locking and unlocking the laser is the main contributor to the measurement uncertainty, and was investigated to be ~ 2 kHz. The accuracy of the H-maser of the National Metrology Institute of Japan (NMIJ) was calibrated by the Coordinated Universal Time (UTC) and confirmed to be better than $1\times10^{-14}$ during the measurement.

There is another frequency uncertainty factor and this results from the frequency shift caused by possible cell contamination. An international investigation shows that this could add an uncertainty of 5 kHz to measurement results [13]. In this measurement, we used an iodine cell that had been used in international comparisons and had shown an agreement of better than 2 kHz with cells from other institutes.

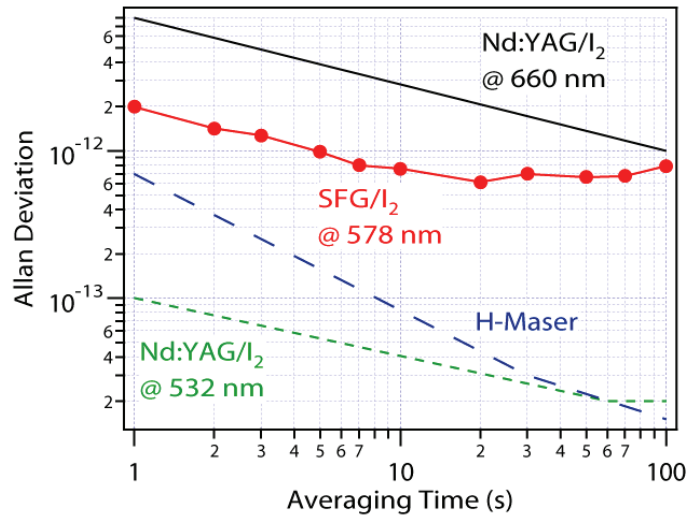

Fig. 5. Allan deviation of the measured beat frequency between the SFG light locked on the $R(37)16-1:a_1$ transition and the comb locked on an H-maser. For comparison, the Allan deviations of an H-maser (dashed curve), an iodine-stabilized Nd:YAG laser at 532 nm (dotted curve), and an iodine-stabilized Nd:YAG laser at 660 nm (solid line) are also shown.

## 4. Discussion and conclusion

A PPLN-WG is a very attractive device for various applications because of its high nonlinear conversion efficiency and capacity for accepting a fiber pigtailing input. With SFG and DFG applications in particular, the conventional method of using an enhancement cavity requires the double resonance of both input lights to achieve high conversion efficiency. This technical difficulty can be overcome by using a PPLN-WG device. As other examples of its use, 1064-nm and 1319-nm Nd:YAG lasers are used for SFG at 589 nm, where sodium D lines are available. On the other hand, a 1064-nm Nd:YAG laser and a 1.5-μm laser diode are used for DFG at 3 μm, where rich absorption lines are available with a number of molecules including $CH_4$. It is worth noting here that the recently developed ridge PPLN WG [27, 28] has not only a higher conversion efficiency but also a higher damage threshold and better long-term reliability than a proton-exchanged PPLN WG [29].

In the present experiment, a 1030-nm Yb fiber laser was used as one light source for SFG. This light source can be easily replaced by a 1030-nm Yb:YAG laser. The monolithic-cavity

Yb:YAG laser has a linewidth of ~ kHz, while the Yb fiber laser has a linewidth of ~ 10 kHz. Frequency jitter is also smaller with the Yb:YAG laser than with the Yb fiber laser. The Yb fiber laser has a wider tuning range of 90 GHz than that of the Yb:YAG laser (40 GHz). Fortunately, the tuning range of the Yb:YAG laser makes it possible to realize SFG with the 1319-nm Nd:YAG laser for applications with both the Yb clock transition and the nearby iodine transitions.

High-resolution spectroscopy of molecular iodine has been employed at 515, 532, 543, 612, 633, 640, and 660 nm with various laser sources. However, there is no iodine frequency reference at the yellow wavelength. The measured absolute frequency of the iodine hyperfine transition serves as an excellent reference at this wavelength. In fact, the measured absolute frequency has been used as a frequency reference by a group at Kyoto University for their Yb project [30].

The frequency measurement of all 21 hyperfine components should lead to a precision measurement of hyperfine splitting (frequency intervals between hyperfine components) of the transition. In terms of molecular physics, it is attractive that accurate hyperfine constants can be obtained from the theoretical fit of precisely measured hyperfine splitting. With accurate measurements of various hyperfine constants, we should be able to derive or improve the formulas for hyperfine constants [31-35]. Research along this line is under way.

In summary, we have developed a stable light source using SFG for high-resolution spectroscopy at 578 nm. The hyperfine transitions of molecular iodine have been observed and used to stabilize the SFG light source. The frequency stability of the SFG light source has reached $2\times10^{-12}$ for a 1-s averaging time. The absolute frequency the R(37)16-1:$a_1$ hyperfine transition is found to be 518304551833 (2) kHz. This transition serves as an excellent frequency reference in the yellow wavelength region.

**Acknowledgments**

We are grateful to M. Imae, Y. Fujii and T. Suzuyama for maintaining UTC at NMIJ.